\documentclass[fleqn, usenatbib]{mnras}

\usepackage{newtxtext,newtxmath}

\usepackage[T1]{fontenc}

\DeclareRobustCommand{\VAN}[3]{#2}
\let\VANthebibliography\thebibliography
\def\thebibliography{\DeclareRobustCommand{\VAN}[3]{##3}\VANthebibliography}

\usepackage{graphicx}	
\usepackage{amsmath}	
\usepackage{amssymb}	
\usepackage{xcolor}
\usepackage{comment}
\usepackage{soul}
\usepackage{float}

\newcommand{\Msun}{M_\odot}
\newcommand{\Mdot}{\dot{M}}
\newcommand{\Mdotstar}{\dot{M}_\ast}
\newcommand{\Mdotin}{\dot{M}_\mathrm{in}}

\newcommand{\Mdotetacrit}{\dot{M}_\mathrm{\eta,touch}}
\newcommand{\Mdotksicrit}{\dot{M}_\mathrm{\xi,touch}}

\newcommand{\MdotLMXB}{\Tilde{\dot{M}}_\mathrm{LMXB}}
\newcommand{\Mdottilde}{\Tilde{\dot{M}}_\mathrm{in}}
\newcommand{\BLMXB}{\Tilde{B}_\mathrm{LMXB}}
\newcommand{\BRMSP}{\Tilde{B}_\mathrm{RMSP}}
\newcommand{\Btilde}{\Tilde{B}}
\newcommand{\PLMXB}{\langle P_\mathrm{LMXB} \rangle}
\newcommand{\PRMSP}{\langle P_\mathrm{RMSP} \rangle}

\newcommand{\Pdot}{\dot{P}}

\newcommand{\Pmin}{P_\mathrm{min}}

\newcommand{\Peq}{P_\mathrm{eq}}
\newcommand{\Porb}{P_\mathrm{orb}}
\newcommand{\Pbreakup}{P_{\rm break-up}}
\newcommand{\Ptd}{P_{\rm touch}}
\newcommand{\Pfinal}{P_{\rm final}}

\newcommand{\rin}{r_\mathrm{in}}
\newcommand{\rlc}{r_\mathrm{LC}}
\newcommand{\rco}{r_\mathrm{co}}
\newcommand{\rone}{r_\mathrm{1}}
\newcommand{\rinmax}{r_\mathrm{in,max}}
\newcommand{\rA}{r_\mathrm{A}}
\newcommand{\reta}{r_\eta}
\newcommand{\rxi}{r_\xi}

\newcommand{\Reta}{R_\eta}
\newcommand{\Rxi}{R_\xi}
\newcommand{\Rinmax}{R_\mathrm{in,max}}
\newcommand{\DeltaR}{\Delta r/r}
\newcommand{\Rstar}{R_{*}}

\newcommand{\OmegaK}{\Omega_\mathrm{K}}

\newcommand{\Lx}{L_\mathrm{X}}

\newcommand{\Lxq}{L_\mathrm{X,quiescence}}

\newcommand{\tint}{\tau_\mathrm{int}}

\newcommand{\gpers}{g~s$^{-1}$}  
\newcommand{\ergpers}{erg~s$^{-1}$}
\newcommand{\ergperscm}{erg~s$^{-1}$~cm$^{-2}$} 

\newcommand{\spers}{s~s$^{-1}$}

\newcommand{\Alfven}{Alfv$\acute{\mathrm{e}}$n~}
\newcommand{\Fx}{F_\mathrm{X}}
\newcommand{\Fav}{F_\mathrm{x,av}}
\newcommand{\x}{\times}

\title[On the lack of X-ray pulsation in most LMXBs]{On the lack of X-ray pulsation in most neutron star low-mass X-ray binaries}

\author[Niang et al.]{
N. Niang,$^{1}$\thanks{E-mail: niang@sabanciuniv.edu}
\"{U}. Ertan,$^{1}$
A. A. Gen\c{c}ali$^{1}$
O. Toyran$^{2}$,
A. Ulubay$^{3}$,
E. Devlen$^{4}$,
M. A. Alpar$^{1}$,
E. G\"{u}gercino\u{g}lu$^{5}$
\\
\\
$^{1}$Sabanc{\i} University, Orhanl{\i}, Tuzla, 34956, \.{I}stanbul, Turkey\\
$^{2}$Merdivenköy mah., Hızır Reis sok., Kad{\i}köy, 34732, \.{I}stanbul, Turkey\\ 
$^{3}$Faculty of Science, Department of Physics, Istanbul University, 34134, Vezneciler, \.{I}stanbul, Turkey\\
$^{4}$Faculty of Science, Department of Astronomy and Space Sciences, Ege University, 35100, Bornova, \.{I}zmir, Turkey\\
$^{5}$National Astronomical Observatories, Chinese Academy of Sciences, 20A Datun Road, Chaoyang District, Beijing 100101, China
}

\date{Accepted 2024 June 24. Received 2024 June 21; in original form 2024 March 18}

\pubyear{2024}

\begin{document}
\label{firstpage}
\pagerange{\pageref{firstpage}--\pageref{lastpage}}
\maketitle

\begin{abstract}
We have investigated whether the lack of X-ray pulsations from most neutron star (NS) low-mass X-ray binaries (LMXBs) could be due to the extension of their inner disc to the NS surface. To estimate the inner disc radii, we have employed the model, recently proposed to account for the torque reversals of LMXBs. In this model, the inner disc radius depends on the spin period as well as the dipole moment and the mass inflow rate of the disc. Our model results indicate that most LMXBs have mass accretion rates above the minimum critical rates required for the inner disc to reach down to the NS surface and thereby quench the pulsed X-ray emission. For most sources X-ray pulsations are allowed when the period decreases below a certain critical value. For the same parameters, the model is also consistent with the observed X-ray luminosity ranges of the individual accreting millisecond X-ray pulsars (AMXPs). The paucity of AMXPs compared to the majority population of non-pulsing LMXBs is explained, as well as the fact that AMXPs are transient sources.
\end{abstract}




\section{Introduction}
\label{Introduction}

Low-mass  X-ray binaries  (LMXBs) harbor a compact object, either a neutron star (NS) or a black hole, with a low-mass 
($M < 1 M_{\odot}$) companion. In these systems, the compact star accretes matter from the donor star through Roche-lobe overflow \citep{FranKingRaine2002}. Since NSs, unlike black holes, have rigid surfaces, the detection of thermonuclear bursts or X-ray pulsations indicate that the binary is hosting a NS. Hereafter we use "LMXBs" to denote LMXBs with NSs. In LMXBs, the matter flowing from the companion forms an accretion disc around the NS. For a given mass inflow rate, $\Mdotin$, the inner disc is cut at a radius, $\rin$, depending on the magnetic dipole moment, $\mu$, of the NS. If the field is sufficiently strong, the inner disc can be truncated at a radius greater than the radius of the NS, $\Rstar$. Velocity of the closed field lines rotating with the star is equal to the Kepler velocity of the disc matter at the co-rotation radius, $\rco$. When $\rin \leq \rco$ , the inflowing disc matter can be channeled by the closed field lines to the magnetic poles of the star. In this case, if the dipole field and the spin axes are not aligned, the emission from the polar caps heated by accretion could be observed as X-ray pulses. For sufficiently high $\Mdotin$ and/or weak dipole field, the inner disc could extend down to the surface of the star. In this case, the star is not likely to produce pulsed X-ray emission. 

Despite extensive observational searches, no pulsations were detected from most of LMXBs \citep[][see \citet{Patruno2021} for a detailed discussion]{Vaughan1994Searches2,Dib2004An1820-30,Messenger2015AX-1,Patruno2018,Galaudage2021DeepSearches}. Among $\sim 190$ known LMXBs, only $40$ sources show coherent X-ray pulsations produced by the mass accretion on to the star \citep[][for details, see Section~\ref{persistent LMXB properties}]{Liu2007, Patruno2021, Avakyan2023}. What could be the physical reason that prevents the pulsed X-ray emission of these sources?

There are several ideas proposed to account for this: (i) NS magnetic fields may not be strong enough to channel the accretion to the magnetic poles \citep{Psaltis1999,Kulkarni2008Rayleigh-Taylor-UnstableSimulations,Romanova2008UnstableSimulations}. (ii) Magnetic fields could be screened under the surface due to long-lasting high mass accretion rates resulting in weak surface magnetic fields \citep{BisnovatyiKogan1974PulsarsSystems,Romani1990AFields,Cumming2001}. (iii) Pulsations could be smeared out while the signal is passing through a hot corona \citep[][but see also \citep{Göğüş2007}]{Brainerd1987,Kylafis1989,Titarchuk2002WhyBinaries,Titarchuk2007CorrelationsX2}. (iv) Gravitational lensing of the X-ray emission could decrease the pulse fraction below the detection limits \citep{Wood1988TheOscillations,Meszaros1988,Ozel2009}. Additionally, non-pulsating sources have higher long-term accretion rates, and the resultant increase in the mass of the star could lead to a more efficient gravitational bending, decreasing the pulse fractions of the signals further \citep{Ozel2009}. (v) Magnetic and rotational axes of the NS could be aligned \citep{Ruderman1991NeutronBinaries,Lamb2009AREGIONS}. (vi) Unstable accretion on to the surface of the NS might prevent the formation of regular pulses when the misalignment between the magnetic field and the spin axis of the NS $< 30^{\circ}$ \citep{Romanova2008UnstableSimulations}. (vii) The binary orbital motion could smear the already weak signals \citep{Bahar2021}. (viii) The presence of complex field structures disrupting the channeled accretion to the poles was also proposed to be the reason for the lack of pulsations \citep{das2022}.

Is it possible that $\rin \simeq \Rstar$ for non-pulsating sources, while $\rin > \Rstar$ for pulsating LMXBs?   This is not likely  for the conventional models assuming $\rin$  close to the \Alfven radius  $\rA = \Big[ \umu^4 / (GM \Mdotin^2) \Big] ^{1/7}$ \citep{davidson1973,lamb1973} where $M$ is the mass of the NS  and $G$ is the gravitational constant, because  $\rA > \Rstar$  for typical LMXBs  (see Section~\ref{persistent LMXB properties}).
In this work, we will test whether the $\rin \simeq \Rstar$ condition is satisfied for most LMXBs using the model recently proposed by \cite{ertan2021}. In this model, $\rin$ is estimated to be much smaller than $\rA$ in both the spin-up and spin-down regimes.

The mass accretion rate on to the star, $\Mdotstar$, can be estimated from the X-ray luminosity $\Lx = G M \Mdotstar / \Rstar$. Since the rotational period, $P$, and its time derivative, $\Pdot$, are not known for non-pulsating LMXBs, $\mu$ cannot be estimated from the disc torque models. However, as LMXBs are progenitors of the radio millisecond pulsars (RMSPs), whose dipole magnetic moments $\mu$ and surface magnetic fields $B$ are inferred from $P$ and $\Pdot$ measurements, their dipole magnetic field distribution should be the same as the distribution for RMSPs. Immediately after the discovery of the first radio millisecond pulsar \citep{Backer1982}; it was proposed that the radio millisecond pulsars (RMSPs) were spun up in LMXBs whose dipole field strength $B = \mu / \Rstar^3$ should be in the range of $\sim 10^8 - 10^9$~G \citep{Alpar1982APulsars, Radhakrishnan1982ONPULSAR}. This was predicted to be the field range of the RMSPs before the first RMSP $\Pdot$ was measured to provide an inferred $B$ value in the predicted range. This process is called the {\em recycling scenario} \citep[for reviews see][]{Alpar2008, Srinivasan2010RecycledPulsars}. The first observational evidence directly supporting this idea came in 1998 with the discovery of the first accretion-powered millisecond X-ray pulsar (AMXP) SAX 1808.4-3658 with $P = 2.5$~ms \citep{Wijnands1998TheJ1808.43658}. In the last two decades, $23$ AMXPs have been discovered with periods less than $10$~ms, and with a sharp cut-off at $1.6$~ms \citep{DiSalvo2020AccretionPulsars, Patruno2021}. The latest evidence for the recycling scenario has been the discovery of three transitional millisecond pulsars (tMSPs) which show transitions between LMXB and RMSP states \citep{Archibald2009ALink, Papitto2013SwingsPulsar, Bassa2014AXSSJ122704859, Jaodand2016TIMINGSTATEb, PapittodeMartino2020}. Evidently, AMXPs are the sources on the evolutionary track leading to RMSPs.
 
How and why are the dipole magnetic fields of the old NS populations of LMXBs and RMSPs so much smaller than the fields of young NSs? It has been proposed that the initial strong dipole fields ($B \sim 10^{12}$~G) of young NSs can decrease by a few orders of magnitude during the initial epoch of accretion from the companion's wind in a high-mass X-ray binary: Spin-down of the NS by the wind accretion decreases the super-fluid vortex density, which enforces a decrease in the density of flux lines in the proton superconductor and thereby reduces the magnetic flux throughout the super-fluid core of the NS \citep[][see the review by \citet{Srinivasan2010RecycledPulsars}]{Srinivasan1989, Srinivasan1990, Sauls1989}. The NS enters the subsequent LMXB epoch of accretion at high rates when the companion fills its Roche-lobe. By this time the magnetic flux in the super-fluid {\em core} has already been reduced to a level corresponding  to an average $\langle B \rangle \sim 10^8 - 10^9$~G. In the early LMXB epoch, the magnetic field in the {\em crust} decreases by diffusion, while the crust is being heated by accretion \citep{Blondin1986IsStar,geppert1994,urpin1995,konar1997}. The accretion process also pushes the current-carrying regions to deeper, high-density layers of the crust with higher conductivity. This process is expected to freeze the eventual surface 
field at $B \sim 10^{8} - 10^{9}$~G and introduce a correlation between the frozen surface field $B$ and the long-term $\Mdotin$, as was shown through numerical simulations \citep{konar1997, Konar2017}. This correlation results in a minimum value, $\Pmin \sog 1$~ms, of the final equilibrium period of the RMSP spun-up by accretion \citep{ertan2021minperiod}. Other explanations for the lack of observed sources with $P \sol 1.4$~ms involve spin-down by gravitational radiation \citep{Patruno2012b,Mahmoodifar2013}.
This picture suggests that the $B$ distribution of LMXBs is likely to be similar to that of RMSPs. Among $\simeq 280$ RMSPs with estimated $B$ and 
$P < 10$~ms, $\simeq 70$ sources are isolated and the remaining are still in binary systems \citep[ATNF Pulsar catalogue, version 1.71,][]{Manchester2005Thecatalogue}\footnote{\url{https://www.atnf.csiro.au/research/pulsar/psrcat/}}. Hereafter we use "RMSPs" to denote the sources with 
$P < 10$~ms. The inferred surface dipole fields of RMSPs have a median value $\BRMSP \simeq 2\times 10^{8}$~G while the average spin period $\PRMSP \simeq 4$~ms. 

In the conventional models, a disc threaded by the  magnetic dipole field of NS is assumed, and the total magnetic torque is calculated by integrating the magnetic torques across the disc within the light-cylinder radius, $\rlc= c/\Omega_{*} $, where c is the speed of light and $\Omega_{*}$ is the angular speed of the star \citep{Ghosh1979AccretionSources, Arons1993MagneticPulsars, Ostriker1995MagnetocentrifugallyZone}. A series of theoretical studies and numerical simulations show that this configuration cannot be sustained, since the diffusion time-scale of the field lines is much longer than the dynamical time-scale of the disc, $\tau_\mathrm{dyn} \sim \OmegaK^{-1}$, where $\OmegaK$ is the Keplerian angular speed \citep{aly1985,wang1987, Lovelace1995Spin-up/spin-downOutflows, Hayashi1996X-Raydisk, Miller1997MagnetohydrodynamicInteraction, Uzdensky2002MagneticallyMagnetosphere, Uzdensky2004Magneticdisks, Fromang2009TurbulentInstability}.

Detailed simulations suggest that the magnetic field lines interact with the inner disc in a narrow boundary layer 
\citep{Lovelace1995Spin-up/spin-downOutflows, Hayashi1996X-Raydisk, Miller1997MagnetohydrodynamicInteraction}. The closed field lines interacting with the matter, inflate and open up within a short interaction time-scale $\tau_\mathrm{int}= |\Omega_{*} - \OmegaK(\rin)|^{-1}$. In the propeller phase, the matter in the boundary layer can be thrown out along the open field lines. Subsequently, the field lines reconnect on a time-scale $\tau_\mathrm{dyn}$ \citep{Lovelace1999MagneticOutflows, Ustyugova2006PropellerStars}. This interaction boundary is at the innermost region of the disc with a narrow radial extension ($\Delta r < r$). Inside $\rin$, the field lines are closed and rotate together with the matter, forming the magnetosphere. Using the basic results of the model proposed by \cite{Lovelace1999MagneticOutflows}, \cite{ertan2017} proposed a model to calculate $\rin$ in the propeller phase, and, the critical condition for the onset of the propeller mechanism. The model was later extended to cover all the rotational phases and the critical conditions for the transitions between these phases \citep{ertan2021} and was utilized to account for the torque reversals of 4U~1626--67 \citep{Gencali2022The162667}. We will use the same model to investigate whether the inner disc could quench the channeled  accretion on to the poles of the star for most LMXBs. In Section~\ref{model}, we briefly describe the model. The properties of the sources  used in our analyses are summarized in Section~\ref{source properties}. We discuss the results of the model calculations in Section~\ref{results&discussion}, and summarize our conclusions in Section~\ref{conclusions}.

\section{The Model}
\label{model}

\begin{figure}
	\includegraphics[width=1\columnwidth]{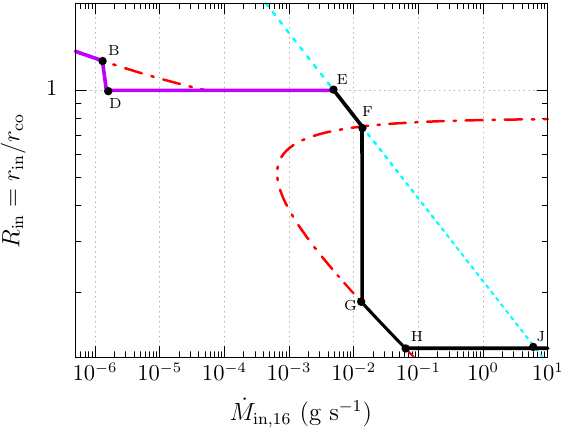}
    \centering
    \caption{The variation of $\rin$ with $\Mdotin$ in the SP, WP (solid purple curve) and SU phases (solid black curve). For this illustrative source, the model parameters are  
    $\DeltaR = 0.2$, $\eta = 0.8$, $\xi = 0.5$, $B = \mu/\Rstar^{3} = 1 \times 10^8$~G, and $P = 10$~ms (see the text for details). The red dot-dashed curves and turquoise dashed line represent $\Reta (\Mdotin)$ and $\Rxi (\Mdotin)$ respectively.}
    \label{fig:case1}
\end{figure}

Here, we describe the model briefly  \citep[for details see][]{ertan2021}. There are three basic rotational phases: strong propeller (SP; spin-down with no accretion), weak propeller (WP; spin-down with accretion), and spin-up (SU; spin-up with accretion). At low $\Mdotin$ levels, the system is in the SP phase with $\rin > \rone = 1.26~\rco$, and all the in-flowing matter is thrown out of the system from the inner disc boundary. This can happen at radii where magnetic field lines compel the matter into synchronized rotation with the NS, and expel it from the system with speeds greater than the escape speed. The maximum value of the inner disc radius where the condition for the strong propeller effect is achieved can be written as:
\begin{equation}	
	\Rinmax^{25/8}~|1 - \Rinmax^{-3/2}| \simeq 0.22~\alpha_{-1}^{2/5}~M_{1.4}^{-7/6}~\Mdot_\mathrm{in,16}^{-7/20}~\umu_{26}~P^{-13/12}_\mathrm{ms}
\label{eq:Rin} 	
\end{equation}
where $\Rinmax=\rinmax/\rco$, $M_{1.4} = (M / 1.4 \Msun )$, $\Mdot_\mathrm{in,16} = \Mdotin / (10^{16}$~\gpers), $\umu_{26} = \umu / (10^{26}$~G~cm$^3$), $\alpha_{-1} = (\alpha /0.1)$ 
is the kinematic viscosity parameter \citep{Shakura1973BlackAppearances}, and $P_\mathrm{ms}$ is the period in milliseconds \citep{ertan2017}. In this phase, the inner disc radius can be written as $\rin = \reta = \eta \rinmax$ with $\eta\lesssim 1$ likely to be close to unity due to the sharp $r$ dependence of the magnetic torque. We also define $\rxi=\xi\rA$ with $\xi \sim 0.5 - 1$ \citep{Ghosh1979AccretionSources, Kluzniak2007}. The solid curve in Fig.~\ref{fig:case1} shows  $\rin$ as a function of $\Mdotin$ for an illustrative source with $B = 10^{8}$~G and $P = 10$ ms. For $\Mdotin$ values smaller than the critical rate corresponding to $\rin = r_{1}$ (point B), the star is in the SP phase. The blue dashed curve shows the variations of $\rxi$ for comparison. It is seen that  $\rin$ is significantly smaller than $\rxi$ in the SP phase. With increasing $\Mdotin$, $\rin$ approaches $\rone$. At times when $\rone \geq \rin > \rco$, the matter expelled from $\rin$ returns back to the disc, and the resultant pile-up moves the inner disc inwards until $\rin=\rco$ (the transition from point B to point D). Accretion on to the star becomes possible at point D and the system goes into the WP phase with continuing spin-down of the star. The WP phase with $\rin=\rco$ can persist for a large range of $\Mdotin$ (from D to E) while $\rxi>\rco$.

This is due to the long $\tint$ in the boundary and the dominance of  magnetic stresses over the viscous stresses, which prevent $\rin$ from penetrating $\rco$. In this phase, all the matter coming to the inner boundary couples to the field lines at $\rco$ and flows along the field lines to the NS.  At the end of the WP phase, the inner disc can penetrate inside $\rco$ when the mass in-flow rate $\Mdotin$ is above a critical level so that the viscous torques dominate the magnetic torques at $\rco$. This approximately corresponds to $\rxi=\rco$ (point E). With further increase in $\Mdotin$ the inner disc enters inside $\rco$ and the SU phase starts. 

\begin{figure}
	\includegraphics[width=1\columnwidth]{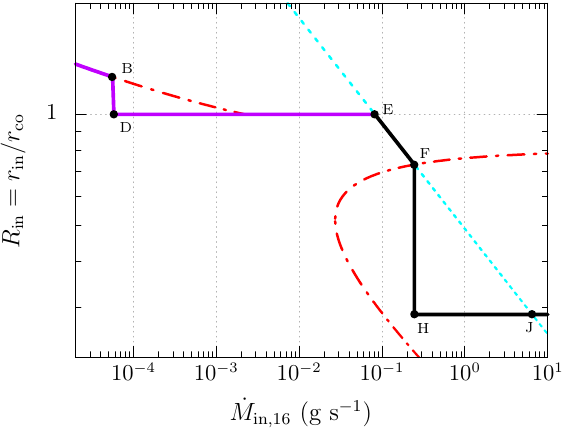}
    \centering
    \caption{Same as Fig.~\ref{fig:case1}, but with $P=3$~ms.}
    \label{fig:case2}
\end{figure}

From the WP/SU transition up to the $\Mdotin$ level at which $\rxi$ crosses the $\reta$ curve (point F), $\rin$ tracks $\rxi$. In the SU phase, $\reta$ is double valued. The upper branch of this solution with a positive slope is unstable. From point F on the unstable branch of $\reta$, the inner disc moves inwards opening the closed field lines to the radius on the stable (lower) branch of $\reta$ corresponding to the same $\Mdotin$ (point G). For $\Mdotin$ beyond this level, $\rin$ decreases tracking $\reta$ until the disc reaches the surface of the star, at point H in Fig.~\ref{fig:case1}. Depending on $B$ and $P$ values, it could be the case that $\reta$ at point G could be smaller than $\Rstar$. In this case, inward propagation of the inner disc terminates at the surface of the star, which is illustrated in Fig.~\ref{fig:case2}. For comparison, we also define $\Mdotin (\rxi = \Rstar) \equiv \Mdotksicrit$, the mass in-flow rate at which the disc inner radius would reach down to the NS surface if it were determined by the \Alfven radius. This is depicted with point J in Fig.~\ref{fig:case1}. It is seen in Figs~\ref{fig:case1} and \ref{fig:case2} that $\rin$ reaches the surface of the star at an $\Mdotin$ much lower than the $\Mdotksicrit$ rate corresponding to $\rxi = \Rstar$, the point where the dashed line depicting $\rxi (\Mdotin)$ intersects $r = \Rstar$. In most cases, $\rin$ estimated in our model is found to be significantly smaller than $\rxi$ in both SP and SU phases. Only for a narrow $\Mdotin$ range,  $\rin$ tracks $\rxi$  during the transition from the WP to the SU phase (between E and F). The inner disc reaches the surface of the star ($\rin = \Rstar$) at a critical $\Mdotin = \Mdotetacrit$. Pulsed X-ray emission is not allowed when $\rin = \Rstar$, which requires $\Mdotin \geq \Mdotetacrit$. 

\begin{figure}
	\includegraphics[width=\columnwidth]{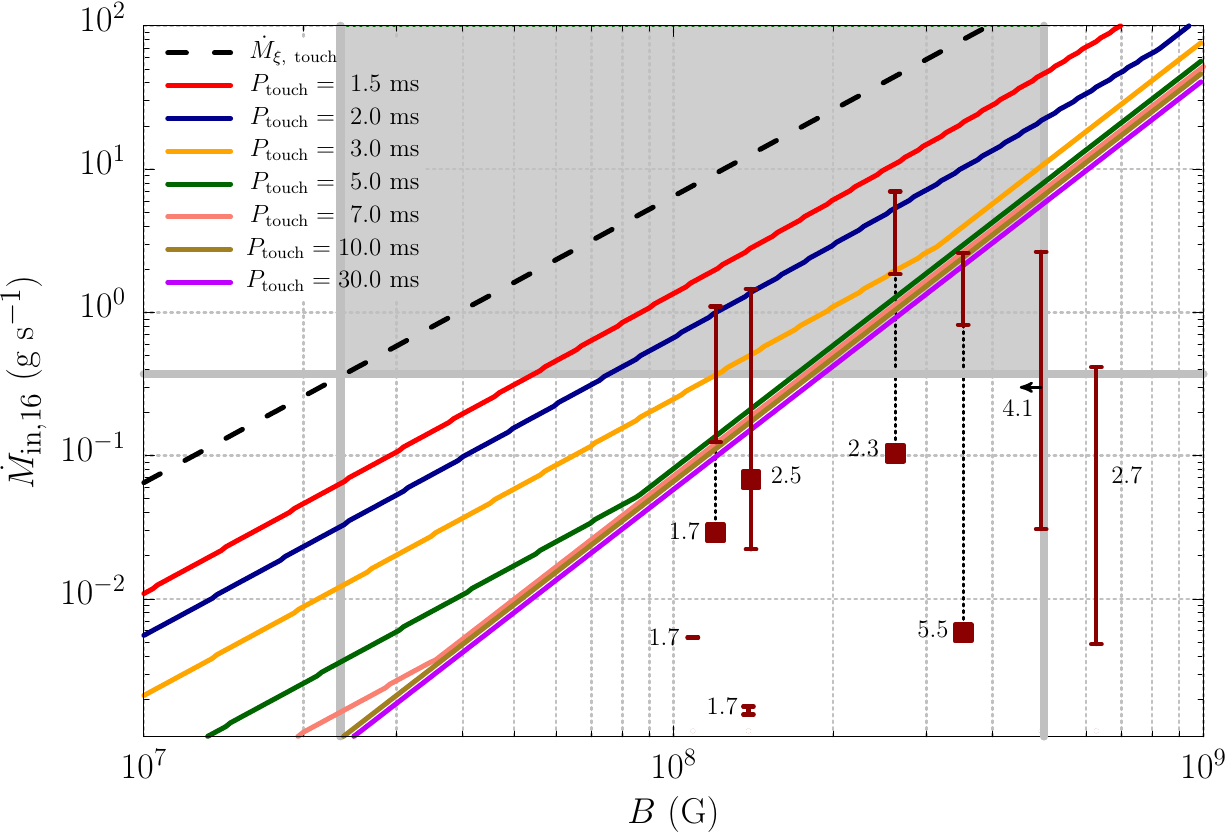}
    \centering
\caption{Illustrative $\Ptd$ curves for different $P$ values (given in the top left panel). On the $\Mdotksicrit$ curve (dashed line), $\rxi = \Rstar$. The model curves are obtained with $\DeltaR = 0.2$, $\xi=0.5$, $\eta=0.8$. For $\simeq 90 \%$ of the non-pulsating persistent LMXBs, $\Mdotstar \gtrsim 4 \times 10^{15}$~\gpers. The $B$ values estimated for $\simeq 90 \%$ of the RMSPs are in the range $2.35 \times 10^7 < B < 5 \times 10^8$~G; this should also be the $B$ range for persistent LMXBs and RMSPs. The region with $\Mdotstar \gtrsim 4 \times 10^{15}$~\gpers~and $2.35 \times 10^7 < B < 5 \times 10^8$~G is shaded in grey. The filled squares indicate the long-term average $\Mdotin$ values (outburst + quiescent phase) obtained by \citet{Watts2008DetectingStars} for $4$ AMXPs. The numbers show the periods of the source in ms. The $B$ values for $6$ AMXPs were obtained assuming a dipole radiation torque with the spin-down rates inferred from long-term period changes. An arrow indicates the upper-limit $B$ value for the source with $P = 4.1$~ms. Solid bars indicates the range of $\Mdotin$ with X-ray pulsations of the sources (see Table~\ref{pulsating sources}).}
   \label{fig:model1}
\end{figure}

\begin{figure}
	\includegraphics[width=\columnwidth]{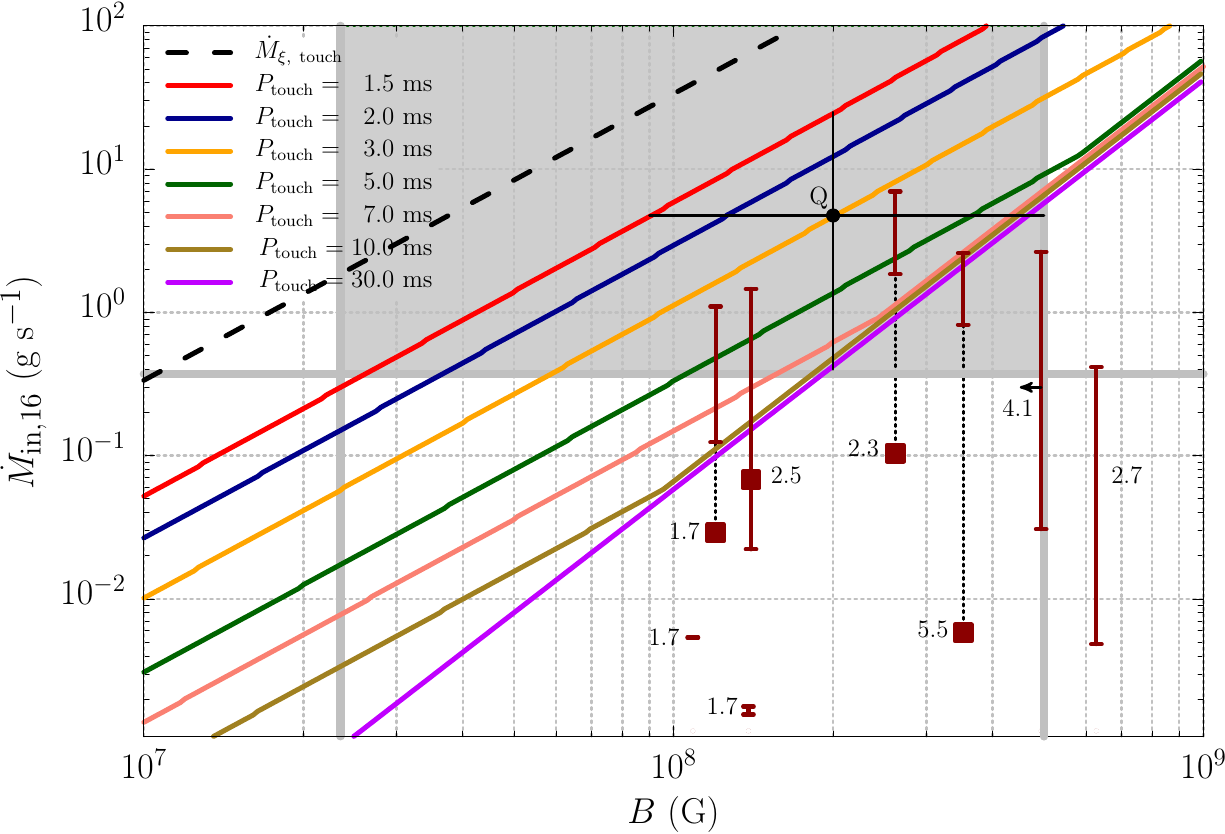}
    \centering
    \caption{The same as Fig.~\ref{fig:model1}, but with $\xi = 0.8$. The median values $\MdotLMXB$ and $\BRMSP = \BLMXB$ intersect at point Q.}
    \label{fig:model2}
\end{figure}

In our model, in addition to $\Mdotin$ and $B$, $\rin$ also depends on $P$, while $\rxi$ has no $P$ dependence. For given $\Mdotin$ and $B$, there is a certain $P$ value at which the inner disc just touches the surface of the NS. We denote this period as the touch-down period, $\Ptd$. For $P \leq \Ptd$, $\rin \geq \Rstar$ and the pulsed X-ray emission is allowed. When  $P > \Ptd$,  the $\Mdotin$ level is above the minimum critical rate required to bring the inner disc to the surface of the star, that is, $\Mdotin \geq \Mdotetacrit$. The value of $\Ptd$ is a function of $\Mdotin$ and $B$. In Fig.~\ref{fig:case1} plotted for $P = 10$~ms and $B = 10^8$~G, $\Mdotetacrit \simeq 6 \times 10^{14}$~ \gpers~(corresponding to point H), therefore the critical period required for $\rin$ to reach the NS surface is $\Ptd \simeq 10$~ms for $(B, \Mdotin) = (10^8~{\rm G}, 6 \times 10^{14}$~ \gpers). In Fig.~\ref{fig:case2}, similarly, $\Ptd \simeq 3$~ms is obtained for $(B, \Mdotin) = (10^8~{\rm G}, 2.5 \times 10^{15}$~ \gpers). For each value $K$ of $\Ptd$, the intersection of the $\rin = \Rstar$ surface with the $P = K$ plane in the $(B, \Mdotin, P)$ parameter space, projected to the $B - \Mdotin$ plane, produces the corresponding $\Ptd = K$ curve, as shown in Figs~\ref{fig:model1} - \ref{fig:model3}. Each of the solid curves seen in Figs~\ref{fig:model1} - \ref{fig:model3} is the locus of ($B, \Mdotin$) combinations corresponding to a particular value $K$ of $\Ptd$ given in the upper left corners of the figures.

In Figs~\ref{fig:model1}-\ref{fig:model3}, the two different slopes seen along the 3, 5, and 7 ms lines correspond to the two alternative histories for the inner disc to reach the stellar surface with increasing $\Mdotin$, as illustrated in Figs~\ref{fig:case1} and \ref{fig:case2}. The segments with higher slopes are produced by solutions for which $\rin$  traces  $\reta$ while the inner disc approaches the star (corresponding to the segment G - H illustrated in Fig.~\ref{fig:case1}). For comparison, we also plotted the $\Mdotksicrit$ line corresponding to $\rxi = \Rstar$ (dashed curve). 

For a source to show pulsed X-ray emission, its current $P$ value should be below the $\Ptd$ line corresponding to its current ($\Mdotin, B$) point. For instance, a hypothetical source with $B = 2 \times 10^{8}$~G and $\Mdotin = 10^{16}$~\gpers~lies on the $\Ptd = 3$~ms line for the model parameters given in Fig.~\ref{fig:model1}. This means that the source can be observed as an X-ray pulsar if its $P < 3$~ms. For another source with $B = 2 \times 10^{8}$~G and $P = 3$~ms, pulsed X-ray emission requires $\Mdotin < 10^{16}$~\gpers. Or, for a source with $\Mdotin = 10^{16}$~\gpers~and $P = 3$~ms, $B$ should be greater than $2 \times 10^{8}$~G to yield pulsed X-ray emission.   

The parameter $\xi$ affects both $\Ptd$ and $\Mdotksicrit$ values, since the WP/SU transition takes place when $\rxi \simeq \rco$. In Figs~\ref{fig:model1} and \ref{fig:model2} we show model curves corresponding to $\xi = 0.5$ and $\xi=0.8$  respectively. The other parameters are identical for both models. Comparing these two figures, it is seen that all $\Ptd$ lines shift upwards with a higher value of $\xi$. In our earlier work, with  similar parameters, we obtained reasonable  model results for the torque reversals of 4U~1626--67 \citep{Gencali2022The162667} and LMXB/RMSP transitions of tMSPs \citep{ertan2018}.

For a persistent LMXB, $\Lx$ is likely to decrease gradually during the long-term evolution. This decay becomes sharper during the late phase of evolution due to Roche lobe decoupling of the companion, the so-called decoupling phase. In this phase, it is estimated that the neutron star decouples from the rotational equilibrium as well \citep[see e.g][]{Tauris2012}. Even if a source does not show X-ray pulsations during the long-term evolution, eventually it is likely to become an X-ray pulsar during the decoupling phase as $\Mdotin$ decreases below the $\Mdotin$ required for $\rin = \Rstar$. Those sources with relatively low mean $\Mdotin$ values are more likely to show pulsed X-ray emission for certain ranges of $\Mdotin$  during their X-ray outbursts. On the other hand, LMXBs in late stages of evolution are likely to become transient X-ray source below a certain $\Lx$ ($\Mdotin$) level, which depends on its disc size, because  the entire disc cannot be kept in the stable high viscosity hot state below this critical $\Lx$. Transient behaviour and emergence of pulsations both occur when $\Mdotin$ decreases below a critical level. Indeed, all AMXPs are transient X-ray sources exhibiting X-ray pulses during their outburst states. A theoretical model attempting to explain the X-ray pulsation behavior of LMXBs is expected to be consistent with the properties of both pulsating and non-pulsating sources. 

\section{Source properties}
\label{source properties}

\subsection{ Pulsating LMXBs (AMXPs)}
\label{PulsatingLMXB}

There are $26$ AMXP sources, including three tMSPs which show coherent X-ray pulsations with $P <10$ ms \citep{Patruno2021,PapittodeMartino2020,Marino2022}. We excluded burst oscillation sources from the pulsating LMXB class, as these sources are powered by thermonuclear bursts and do not show coherent X-ray pulsations resulting from the channeled mass flow to the poles \citep{Watts2012}. Some of the AMXPs also show thermonuclear X-ray bursts. There are $\Pdot$ estimations for $6$ AMXPs from their long-term $P$ variations during the X-ray quiescent states. Magnetic dipole moments can be estimated using these $\Pdot$ values with the assumption that magnetic dipole radiation is the dominant torque mechanism in the quiescent state (Table~\ref{pulsating sources}). From the dipole-torque formula, the dipole field strength at the equator of the NS can be estimated as $B \simeq 3.2 \times 10^{19} \sqrt{P \Pdot}$~G.  We summarize these estimates below for the individual sources. We note that, if the disc torque dominates the dipole torque for a particular source, the dipole formula overestimates the actual $B$ value. We will use estimated $\Mdotin$ ranges and $B$ values of these $8$ pulsating sources (Table~\ref{pulsating sources}) together with non-pulsating persistent LMXBs (Table~\ref{nonpulsating}) to test our model predictions. We have excluded the tMSPs from this analysis, since the physical mechanisms producing their X-ray pulses is not clear and may not be channelled accretion onto the NS \citep{Papitto2019,Veledina2019,Baglio2023}.

\subsubsection*{\textbf{Swift J1756.9–2508}}

Swift J1756.9-2508 was discovered in its 2007 outburst with $P = 5.5$~ms \citep{Krimm2007DiscoveryCompanion, Markwardt2007SWIFTPulsar}. The distance is estimated to be $d \simeq 8-8.5$~kpc assuming it is close to the Galactic Center \citep{Krimm2007DiscoveryCompanion}. The long-term spin-down rate is estimated to be $(2.2 \pm{0.8}) \times 10^{-20}$~\spers~from the period difference between the 2007 and 2018 outbursts \citep{Bult2018}. Assuming the source is spun down by the dipole torque during the X-ray quiescent state, we estimate $B \simeq 3.5 \times 10^8$~G.

\subsubsection*{\textbf{IGR J17511–3057}}

IGR J17511-3057 was first detected in its 2009 outburst with $P = 4.1$~ms \citep{Markwardt2009RXTEDetected, Baldovin2009INTEGRALIGRJ17511-3057, Bozzo2015INTEGRALJ17511-3057}. The source was in quiescence from 2009 to 2015. Taking the $P$ difference between the outbursts of 2009 and 2015 \citep{Papitto2016TheXMM-Newton} an upper limit to the long-term $\Pdot$ was estimated to be $5.8 \times 10^{-20}$~\spers. This corresponds to the upper limit $B \lesssim 5 \times 10^8$~G. 

\subsubsection*{\textbf{IGR J17494-3030}} 

IGR J17494-3030  was first discovered during its outburst in 2012  \citep{Boissay2012, Chakrabarty2013}.  The coherent $2.7$~ms pulsations were detected at the next outburst in 2020 \citep{Ng2021}. The distance to the source is not well constrained but estimated to be about $8$~kpc due to its proximity to the Galactic Center \citep{Padilla2013}. Following the 2020 outburst, a recovery of X-ray pulsations from the archival observations of the 2012 outburst allowed the measurement of the secular $\Pdot = (1.45 \pm 0.5) \times 10^{-19}$~\spers~\citep{Ng2021}. This gives $B \simeq 6.3 \times 10^8$~G.

\subsubsection*{\textbf{SAX J1808.4–3658}}

Observation of coherent $2.5$~ms pulsations during the 1998 outburst revealed the source as the first AMXP \citep{Wijnands1998TheJ1808.43658}. Since then, the source underwent $9$ outbursts each lasting $\sim 1$~month \citep{Hartman2008, Sanna2017,Bult2020}. The distance estimated from photospheric radius expansion (PRE) is $3.5 \pm 0.1$~kpc \citep{Galloway2006, Galloway2008}. During some of the outbursts the sign of $\Pdot$ could not be constrained due to timing noise \citep{Hartman2008}. The long-term $\Pdot$ variations measured between 1998 and 2022 give spin-down $\Pdot$ values in the range of $10^{-21} - 10^{-20}$~\spers~\citep{Hartman2009, Patruno2012, Sanna2017, Bult2020}. The $\Pdot$ value estimated from the measurements between 1998 and 2022 \citep{Illiano2023} including the most recent outburst, is $\simeq 7.2 \times 10^{-21}$~\spers, which gives $B \simeq 1.4 \times 10^8$~G. 

\subsubsection*{\textbf{XTE J1751–305}}

XTE J1751-305 was discovered in 2002 with $P = 2.3$~ms during the brightest and longest outburst among 4 observed outbursts \citep{Markwardt2002DiscoveryJ1751305, Papitto2008MeasuringJ1751-305}. From the  $P$ values measured at the end of the 2002 outburst and at the beginning of the 2009 outburst, it was estimated that the source slowed down with an average $\Pdot = (2.9 \pm{0.6}) \times 10^{-20}$~\spers~\citep{Riggio2011SecularJ1751-305}. The effects of the two outbursts between these two measurements on the $\Pdot$ measurement are not likely to be significant due to their short duration (a few days) and low luminosity. From the magnetic dipole torque formula, $B$ is estimated to be $\simeq 2.6 \times 10^8$~G.

\subsubsection*{\textbf{IGR J00291+5934}}

IGR J00291+5934 is the fastest rotating AMXP with $P = 1.67$~ms \citep{Markwardt2007OutburstJ1751-305}. The distance to the source is constrained at $4.2 \pm 0.5$~kpc from PRE during a thermonuclear burst observed in the 2015 outburst \citep{DeFalco2017TheObservations}. The long-term spin-down rate during the quiescence between the 2004 and 2008 outbursts is estimated to be $\Pdot \simeq (1.1\pm0.3)\times10^{-20}$~\spers~\citep{Papitto2011SpinPulsar}. This gives $B \simeq 1.4 \times 10^8$~G for spin down with a pure dipole torque. Using the $P$ values measured at the end of the 2004 and at the beginning of the 2015 outbursts \citep{Papitto2011SpinPulsar, DeFalco2017TheObservations}, we estimate $\Pdot = 8.9 \times 10^{-21}$~\spers~assuming that the 2008 outbursts did not spin-up the source significantly. This corresponds to $B \simeq 1.2 \times 10^8$~G.

\subsection{Non-pulsating persistent LMXBs}
\label{persistent LMXB properties}

The non-pulsating LMXBs given in Table~\ref{nonpulsating} are selected from the recently released LMXB catalogue {\em XRBcats}\footnote{\url{http://astro.uni-tuebingen.de/~xrbcat/}} \citep{Avakyan2023}. The catalogue, listing $62$ parameters for each of $349$ sources including black hole systems, has been built upon previous LMXB catalogues \citep{Liu2007, Ritter2003}.  Among the sources catalogued, $179$ are known to be NS systems identified by their X-ray bursts or coherent X-ray pulsations. In addition, $9$ sources are also classified as NSs based on their X-ray spectral properties. Out of these $188$ NS systems, $148$ sources are non-pulsating LMXBs. Some of the non-pulsating sources have relatively stable $\Lx$, while others  show strong variability. Adopting the convention described by \citet{Bahramian2022}, we have included in the persistent LMXB class those systems which do not have quiescent states and/or $\Lx$  variations of more than three orders of magnitude.

For the X-ray faint  LMXBs,  especially transient sources in their quiescent states ($\Lx \sim 10^{31} - 10^{32}$~\ergpers) lasting from years to decades \citep{Bahramian2022,Chaty2022}, the photon count rates are usually too low to analyze the data for the presence of a pulsed component. In addition, there are large uncertainties in the distance estimations. Some  transient or persistent LMXBs in the low $\Lx$ states ($\Lx \lesssim 10^{35}$~\ergpers) could be in the SP phase during which a large fraction of the inflowing disc matter is expelled from the inner disc,  which might also be the reason for the lack of X-ray pulsations. Considering these possibilities, we exclude the transients and X-ray faint persistent sources ($\Lx < 10^{35}$~\ergpers). Excluding also those with distance uncertainties greater than $30 \%$,  we will  include $24$  persistent sources (see Table~\ref{nonpulsating}) in  our analysis of non-pulsating LMXBs. Orbital periods, $\Porb$,  average X-ray flux, $\Fav$,  estimated $d$, $\Mdotin$, and $\Lx$ values of  these sources  are given in  Table~\ref{nonpulsating}.  The source distances were estimated from the distance to the hosting Globular Cluster (GC) for LMXBs in GC, observations of PRE in X-ray bursters \citep{Kuulkers2003}, and astrometry for LMXBs with identifiable counterparts \citep{Arnason2021,Avakyan2023}. 

With reasonable assumptions, we aim to test with our model whether $\rin \simeq \Rstar$ for most LMXBs with parameters that yield results consistent with the properties of AMXPs 
as well. 
We cannot directly estimate the dipole moments of non-pulsating LMXBs, since  we do not know their $P$ and $\Pdot$  values. It is reasonable to assume that their $B$ distribution is similar to that of RMSPs with a median $\BLMXB = \BRMSP \simeq 2\times 10^{8}$~G, since the magnetic fields are likely to decay and freeze in the early stages of the recycling \citep{konar1997, Konar2017}. We calculate the $\Mdotin$ values from the observed X-ray fluxes, $\Fx$,  and the estimated distances using $\Lx = 4 \upi~d^{2}~F_{X}$. We take the average of the minimum and the maximum $\Fx$ values from the LMXB catalogue \textit{XRBcats} \citep{Avakyan2023}. The median $\Lx$ estimated for these sources is around $1\x 10^{37}$~\ergpers~corresponding to a $\MdotLMXB \simeq 5 \x 10^{16}$~\gpers.

\section{Results and Discussion}
\label{results&discussion}

The model curves seen in Figs~\ref{fig:model1} and \ref{fig:model2} are obtained with $\xi = 0.5$ and $\xi = 0.8$ respectively while the other parameters are the same ($\DeltaR = 0.2$ and $\eta=0.8$). The grey area covers about $90 \%$ of the $\Mdotin$ values estimated for non-pulsating persistent LMXBs and $\simeq 90 \%$ of the $B$ values estimated for the RMSPs. For a source on a point along a $\Ptd$ line (solid lines), the $\rin = \Rstar$ condition is satisfied when the current $P$ of the source is greater than that particular $\Ptd$ value. For comparison, dashed lines show $\Mdotksicrit$ curves along which $\rxi = \Rstar$. The value of $\Mdotksicrit$ for a given $B$ and $\Mdotin$ pair is independent of the $P$ value. For the sources below the $\Mdotksicrit$ curve, $\rxi > \Rstar$. For both $\xi = 0.5$ and $\xi = 0.8$, it is seen in Figs~\ref{fig:model1} and \ref{fig:model2}, that $\rxi > \Rstar$ for a large fraction of LMXBs. This means that a model with $\rin \simeq \rxi = \xi \rA$ cannot account for the lack of X-ray pulsations from  a large fraction ($\simeq 75 \%$ ) of LMXBs by the extension of the inner disc to the surface of the NS. Therefore, if the lack of pulsations is due to the disc extending to the NS surface, the inner disc radius is not determined by the \Alfven radius. By contrast, in our model, which takes into account the effects of stellar rotation in determining the inner disc radius, the inner disc does reach down to the NS surface in a wide range of circumstances.

Indeed, $\rin > \Rstar$ and X-ray pulsations are allowed only when the source is located below its $\Ptd$ line in the $B - \Mdotin$ plane, that is to say, below the line with the $\Ptd$ value equal to the current period $P$ of the source. For instance, for the model given in Fig.~\ref{fig:model2}, a source with $P = 3$~ms at present can show X-ray pulsations if it is located below the $\Ptd = 3$~ms line. The vertical bars show the estimated $\Mdotin$ ranges of AMXPs along which they exhibit coherent X-ray pulsations. Their dipole field strengths were estimated from the long-term $P$ variations between the outbursts as described in Section~\ref{PulsatingLMXB}. 
The numbers near the bars show the $P$ values of these AMXPs in units of ms. The squares on or below the bars show the long-term average $\Mdotin$ values estimated by \cite{Watts2008DetectingStars}. It is seen in Fig.~\ref{fig:model1} that the bars of some AMXPs (XTE J1751-305, SAX J1808.4-3658, and SWIFT J1756.9-2508 with periods of $2.3, 2.5$, and $5.5$~ms respectively) extend above their $\Ptd$ levels. For instance, for the $2.5$~ms source, pulsed emission is allowed below $\Mdotin \simeq 6 \times 10^{15}$~\gpers, while its bar extends to above $1 \times 10^{16}$~\gpers. The model curves in Fig.~\ref{fig:model2}~($\xi =0.8$), are in better agreement with the observations: In this case it is seen that the $\Mdotin$ ranges of all AMXPs remain below their individual current $\Ptd$ lines. In particular, the $\Ptd$ level for SAX J1808.4-3658, with $P = 2.5$~ms, shifts to $\simeq 3 \times 10^{16}$~\gpers~while its entire bar now remains below this level. 

\begin{figure}
	\includegraphics[width=\columnwidth]{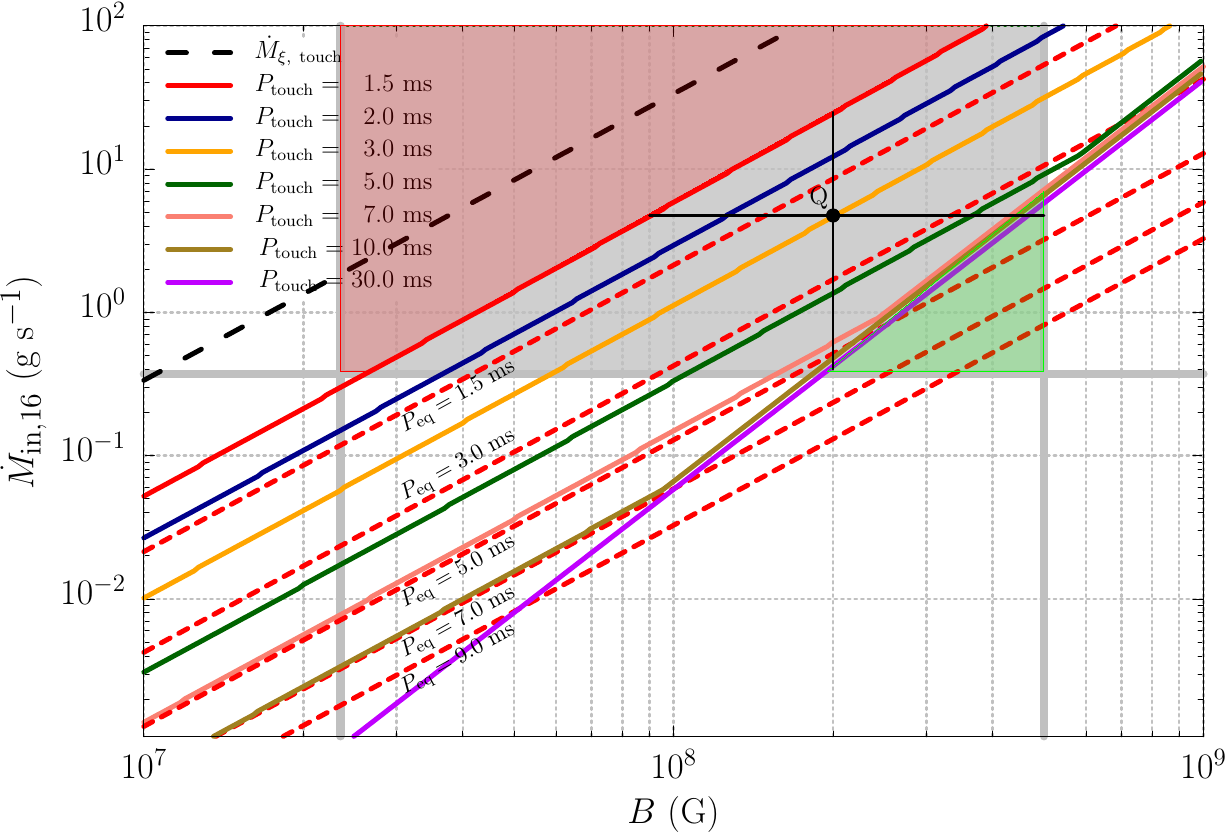}
    \centering
    \caption{The same as Fig.~\ref{fig:model2}. $\Peq$ lines are also plotted with the red dashed lines. In the green area, sources with any $P$ could exhibit X-ray pulses. We estimate that the number of sources in the red region is negligible (see the text for details).}
    \label{fig:model3}
\end{figure}

We now address the properties of the non-pulsating LMXBs with model results for $\xi = 0.8$, a choice consistent with the $\Mdotin$ ranges of all AMXPs. In Figs~\ref{fig:model2} and \ref{fig:model3}, the reference point Q is at the intersection of the lines representing the median values $\BLMXB \simeq \BRMSP  \simeq 2 \times 10^8$~G and $\MdotLMXB$ $\simeq 5 \times 10^{16}$~\gpers~for the sample of non-pulsating persistent LMXB sources listed in Table~\ref{nonpulsating} (see Section~\ref{persistent LMXB properties}). The point Q lies on the $\Ptd = 3$~ms line which means that a source at this point can show X-ray pulsations if its $P < 3$~ms. A low mass X-ray binary would start to emit X-ray pulses when it spins up to a $P$ less than its current $\Ptd$. The long-term gradual $\Mdotin$ decay helps in reaching low enough $\Mdotin$ levels to achieve the 
$P < \Ptd$ condition for the emergence of the source as an AMXP, especially during the late phase of LMXB evolution, when the decrease of $\Mdotin$ becomes sharper in the so-called decoupling phase \citep[See e.g][]{Tauris2012}. When $\Mdotin$ (therefore $\Lx$) decreases below a critical value the disc will not be hot enough to sustain a high viscosity and transport the mass inflow efficiently. The source then starts to exhibit transient behaviour between a high luminosity outburst state and a low luminosity, quiescent state, corresponding to hot and cold disc states. For many sources, the critical $\Mdotin$ for the period to fall below the threshold, $P < \Ptd$ and the source to become an AMXP seems to be smaller than the critical rate below which the source becomes transient, as all the AMXPs known at present are transient sources.

It is reasonable to assume that with age the $P$ distribution of LMXBs is evolving towards that of RMSPs which have an average $\PRMSP \simeq 4$~ms for the sources with $P < 10$~ms \citep[ATNF Pulsar catalogue, version 1.71,][]{Manchester2005Thecatalogue}, while $\PLMXB$, and the period distribution of LMXBs of all ages are of course not known. The $P$ distributions of RMSPs and AMXPs show a sharp cut-off at a minimum, $\Pmin \simeq 1.4$~ms \citep{Hessels2006, Patruno2021}, larger by a factor of $3$ from the critical (break-up) period of NSs, which is estimated to be $\Pbreakup \simeq 0.5$~ms. The observed $\Pmin$ cut-off implies that $\Pmin = \Pbreakup = 0.5$ ms is not reached by LMXBs.
The fact that no sources with periods close to the $\Pbreakup \simeq 0.5$~ms are observed, and theoretical explanations, summarized in Section~\ref{Introduction}, as to why such periods will not be reached, lead us to conclude that the sources with $\rin = \Rstar$ are not in rotational equilibrium: If the disc does reach down to the NS, the rotational equilibrium at the star surface would require $P = P_{\rm K} (\Rstar) \simeq 0.5$~ms. This means that the sources that are spinning up with $\rin = \Rstar$ have $P$ values greater than the maximum $P$ required for pulsations, $\Ptd$. For a given source, $\Ptd$ is longer than the rotational equilibrium period $\Peq$ which is obtained by equating $\rxi$ to $\rco$ as
\begin{equation}	
	P_\mathrm{eq} = 2.09 ~ \xi_{0.8}^{3/2} ~M_{1.4}^{-5/7} ~\umu_{26}^{6/7} ~ \Mdot_\mathrm{in,16}^{-3/7}~{\rm ms}
\label{P_eq} 	
\end{equation}
where $\xi_{0.8} = \xi / 0 .8$. For instance, for a source at  point Q seen in Fig.~\ref{fig:model2}, $\Ptd \simeq 3$~ms while equation~(\ref{P_eq}) gives $\Peq \simeq 1.9 $~ms for this point. $\Peq$ is the minimum period that can be achieved by a source. For a constant $ \Mdotin \simeq 5 \times 10^{16}$~\gpers, the source spins up without X-ray pulsation until $P = \Ptd = 3$~ms. With further decrease in $P$, the inner disc decouples from the surface of the star switching on the X-ray pulses, and the source continues to spin up until $\rin = \rco > \Rstar$ and $P = \Peq$. 

The formation rate of LMXBs is estimated to be $\sim 10^{-6} - 10^{-5}$~yr$^{-1}$ \citep{Van2021LMXB,Shao2015}. While sources in the early stages of LMXB evolution are not likely to pulse in X-rays, sources in later stages are more likely to show X-ray pulsations and become AMXPs. 
For a non-pulsating LMXB located at point Q at present, $\Ptd = 3$~ms and $\Peq \simeq 2$~ms. This source is not likely to emerge as an AMXP at point Q. As $\Mdotin$ decreases, both $\Ptd$ and $\Peq$ will increase. The source will become an AMXP when its period $P$ becomes less than the current $\Ptd$. It might start to show pulsations when $\Mdotin$ has decreased to $10^{16}$~\gpers, for instance. In this case, the source appears as an AMXP with $P = 6$~ms and subsequently spins up till $P = \Peq \simeq 4$~ms, assuming $\Mdotin$ remains constant during this phase.

In the red region above the $\Ptd = 1.5$~ms line in see Fig.~\ref{fig:model3}, which corresponds to $\Peq \simeq 1.0$~ms, the touch-down period $\Ptd < 1.5$~ms, precluding any AMXPs in this region, since $P < \Ptd$ is required for pulsations. Assuming that non-pulsating sources with $\Peq < 1 $~ms, if there are any, are  rare, we expect that the number of sources 
evolving in this region is negligible.  
In the green region in Fig.~\ref{fig:model3}, it is seen that the X-ray pulses are allowed for any $P < 10$~ms. The number of sources (AMXPs) in this region is likely to be small compared to the number of non-pulsating LMXBs. We expect that  most of the LMXBs are located in the grey region in Fig.~\ref{fig:model3}. Comparing  the $\Ptd$ and $\Peq$ lines, we find $\Ptd \simeq (1.5 - 2.5) \times \Peq$ in the grey area. For a particular non-pulsating LMXB, $P$  should be greater than its $\Ptd$ at present, since, otherwise it would emit X-ray pulses. 

For these rough estimates of the distribution of sources in $(B, \Mdotin)$, we have noted that the axes $\Mdotin = \MdotLMXB$ and $B = \Tilde{B}_\mathrm{LMXB, RMSP}$ intersecting at the reference point Q divide the $B - \Mdotin$ plane into four quadrants, each of which contains $25 \%$ of the sources. For sources evolving with each $(B, \Mdotin)$ pair of initial values, assuming that $\Mdotin$ remains constant throughout LMXB evolution, the distribution of $P(t)$ values is affected by the torque law $\dot{\Omega}[\Omega(t)]$ or $\Pdot[P(t)]$. We make the simplifying assumption that the $P$ distribution is uniform for each evolutionary track $(B, \Mdotin)$.

As the individual LMXB sources evolve by spin-up towards the eventual RMSP phase, the mean and median LMXB periods $\PLMXB \simeq \Tilde{P}_{\rm LMXB}$ should  be larger than the mean and median RMSP periods; $\PLMXB > \PRMSP \simeq \Tilde{P}_{\rm RMSP} \simeq 4$~ms \citep[ATNF Pulsar catalogue, version 1.71,][]{Manchester2005Thecatalogue}. For an order of 
magnitude estimation, let us take $\PLMXB \simeq 2~\PRMSP \simeq 8$~ms. In this case, $\PLMXB = 8$~ms represents the median period $\Tilde{P}$ for LMXBs. This means about half of the sources should have $P > 8$~ms and cannot show pulsations inside the grey region where most of the sources are estimated to be located. For the remaining $50 \%$, the sources with $P = 1.5 - 8$~ms, $\Tilde{P} \simeq \langle P \rangle \simeq 5$~ms. If their $\Btilde$ and $\Mdottilde$ also cross near the point Q on the $\Ptd = 3$~ms line, a large majority of the sources in the grey region with $P = 5 - 8$~ms cannot show pulsations either, as they are likely to reside above the $\Ptd = 5$~ms line in the $B - \Mdotin$ plane. Assuming that the sources in this $P$ range are distributed roughly uniformly in the grey area, we estimate that the fraction of the non-pulsating sources in this group is greater than about $90 \%$. For the remaining $\simeq 25 \%$ of the sources with $P = 1.5 - 5$~ms, $\Tilde{P} \simeq \langle P \rangle \simeq 3$~ms, equal to $\Ptd$ at point Q indicating that for about half of them $P < \Tilde{P} \simeq \Ptd$ so that they can show X-ray pulsations. This order of magnitude estimation with the assumption $\PLMXB \simeq 2~\PRMSP$, implies that more than $80 \%$ of LMXBs cannot show pulsations. The exact position of the point Q does not change this result significantly. Our estimations using $P < 10$~ms for RMSPs do not change significantly if $P \sol 16$~ms is adopted to define the RMSP population, as proposed recently \citep{Halder2023} altering the mean period only slightly, to $\PRMSP \simeq 4.4$~ms.  

The mass inflow rate $\Mdotin$ starts to decrease in late stages of binary evolution, especially when the companion star is no longer filling its Roche lobe (the decoupling phase). Finally when $\Mdotin$ from the companion stops altogether, the disc around the NS will quickly become passive on a $\sim 10^5$~yr timescale \citep{ertan2014}, becoming a dust disc, relevant to the possible formation of planets around RMSPs \citep{Wolszczan1992}. The NS now emerges as a radio millisecond pulsar, with some period $\Pfinal$ acquired at the end of LMXB evolution. This is the initial period of the RMSP, which does not change much thereafter as the dipole torque, with the low B values is very ineffective. We can therefore use the observed $P$ distribution of the RMSPs as the distribution of $\Pfinal$ values at the end of LMXB evolution. For about $77 \%$ of the RMSPs in the ATNF catalogue $ P < 5 $~ms, while $5 < P< 7$~ms for $16 \%$, and $7 < P< 10$~ms for only $\simeq 7 \%$ of the sources.

What happens to a NS in the final stages of LMXB evolution? A source entering this decoupling phase will evolve on a constant $B$ line in the $B - \Mdotin$ plane. With average $\Mdotin$ decreasing below $\sim 5 \x 10^{15}$~\gpers, sources are likely to become transient and evolve below the shaded area in Figs~\ref{fig:model1} - \ref{fig:model3} as transient sources. As $\Mdotin$ decreases the corresponding current value of $\Ptd$ increases (the source is evolving downwards in the $B - \Mdotin$ plane, crossing lines with successively larger $\Ptd$ values). When the $\Ptd$ value at the current location in the $B - \Mdotin$ plane becomes larger than the current period $P$, the source will become an AMXP. The source will reach rotational equilibrium when its current period equals the $\Peq$ value at the current location in the $B - \Mdotin$ plane. About $77 \%$ of LMXBs will end with $\Pfinal < 5$~ms, since about $77 \%$ of RMSPs have $P \lesssim 5$ ms. These will have reached the AMXP phase only if their final $(B,\Mdotin)$ values are below the green $\Ptd = 5$~ms line, and they will have reached rotational equilibrium if their final $(B,\Mdotin)$ values reach the $\Peq = 5$~ms line. These conditions are unlikely, especially at the lower $\Mdotin$ values, for weaker $B$ fields. Transient LMXBs with relatively low $B$ and sufficiently long $P > \Ptd(B, \Mdotin) $ may not show X-ray pulsations for large ranges of $\Mdotin$. For instance, for the model given in Figs~\ref{fig:model2} and \ref{fig:model3}, a $10$-ms source with $B = 4 \x 10^{7}$~G (Fig.~\ref{fig:model2}) does not show X-ray pulsations for $\Mdotin > 10^{14}$~\gpers. On the other hand, some LMXBs with relatively strong dipole fields ($\sim 10^{9}$~G) could be in the SP phase at such low $\Mdotin$ levels. These sources do not show X-ray pulsations either. Thus many LMXBs will end their evolution as non-pulsating sources, and without having reached $\Peq$.

All RMSPs and AMXPs have $P \lesssim 10$~ms. Therefore all sources that lie below the $\Ptd = 10$~ms line in Fig.~\ref{fig:model3} have $P < \Ptd$ and have to be AMXPs. This is exemplified by the sources in the green region, IGR J17511-3057 and SWIFT J1756.9-2508, with $P = 4.1$~ms and $P = 5.5$~ms, respectively. For all the AMXPs  with inferred $(B, \Mdotin)$ values, $P < \Peq$: the current periods are less than the current equilibrium periods corresponding to the $(B,\Mdotin)$ ranges inferred for each AMXP, as can be seen from Figs~\ref{fig:model2} and \ref{fig:model3}. These sources seem to have spun up in earlier epochs, at $\Mdotin$ levels higher than their current average $\Mdotin$. This suggests that these sources are currently evolving in the decoupling phase, with decreased $\Mdotin$.

\section{Conclusion}
\label{conclusions}

We have analyzed whether the inner disc could extend down to the surface of the NS, and thereby quench the pulsed X-ray emission preventing channeling of the mass flow to the poles of the star for most LMXBs. This is not in line with models estimating the inner disc radius, $\rin$, to be close to $\rA$, as $\rA$ is found to be  much greater than $\Rstar$ for the majority of LMXBs (see Figs~\ref{fig:model1} - \ref{fig:model3}). For the $\rin$ calculations, we have employed the model recently proposed by \cite{ertan2021} to account for the torque reversals of LMXBs and ongoing accretion at low $\Lx$ levels of tMSPs \citep{ertan2017, ertan2018, Gencali2022The162667}. In this model, $\rin$ depends on the NS period $P$, as well as $\Mdotin$ and $B$. The inner disc radii are found to be much smaller than $\rA$ in both the spin-up (SU) and the strong-propeller (SP) phases ($\rin$ traces $\rA$ only for a narrow $\Mdotin$ range, during the torque reversals close to rotational equilibrium). For given $\Mdotin$ and $B$, there is a critical period $\Ptd$, such that X-ray pulsations are switched on when the current $P$ is shorter than the current value of $\Ptd$, favoring the sources with shorter $P$ values for pulsed X-ray emission. Using the model calculations for $\Ptd(B,\Mdotin)$, we have made statistical estimates employing the $\Mdotin$ distribution for the sample of LMXBs with distance and $\Lx$, the $B$ distribution of RMSPs, and assuming uniform distribution of $P(t)$ values. Our results confirm that most LMXBs must have $\Mdotin$ levels above the minimal values required, at the likely $B$ and $P$ values, to satisfy the $\rin = \Rstar$ condition ($P > \Ptd(B, \Mdotin)$). Our results thus imply that the extension of the inner disc to the surface of the NS is common and is likely to be the main reason for the lack of X-ray pulsations from a large fraction of LMXBs. In the final stages of LMXB evolution, the current value of $\Ptd$ increases as $\Mdotin$ decreases. After $\Mdotin$ decreases below a critical level, sources become transient, with transitions between higher luminosity outburst states and quiescent states. Thus, AMXPs emerge at the final phase of evolution. The model is also consistent with the observed $\Lx$ ranges of those individual AMXPs for which long-term $\Pdot$ measurements, providing estimations of the magnetic dipole field $B$, are available. As a final remark, non-pulsating transient sources with $P > \Ptd$  can show orders of magnitude variation in their $\Lx$ without emitting X-ray pulsations \citep[as observed; see e.g.][]{Avakyan2023, Bahramian2022}.

\section*{Acknowledgements}

We acknowledge research support from Sabanc{\i} University, and from T\"{U}B\.{I}TAK (The Scientific and Technological Research Council of Turkey) through grant 123F083. EG is supported by National Natural Science Foundation of China (NSFC) programme 11988101 under the foreign talents grant QN2023061004L.

\section*{Data Availability}

No new data were analysed in support of this paper.

\begin{table*}
	\centering
	\caption{AMXP sources that are used in our analysis. $B$ values are estimated from dipole torque formula. $\Pdot$ values are estimated from the long-term $P$ variations (see text for details). For AMXPs, the given  $\Lx$ range with X-ray pulsations ($3 - 20$ keV) were taken from \citet{Mukherjee2015ThePulsars}. }
	\label{pulsating sources}
	\begin{tabular}{ccccccccc} 
		\hline
		
		{\bf AMXPs} & $P$ & Outbursts & Pulsed $\Lx$ interval & $\Lxq$ & Distance & $\Pdot$ & $B$ & Ref. \\
							
        & (ms) &  & ($10^{36}$~\ergpers) & ($10^{32}$~\ergpers) & (kpc) & ($10^{-20}$~\spers) & ( $10^{8}$~G) &\\						
        \hline					
							
        SWIFT J1756.9-2508 & $5.5$ & $2007, 2009, 2018$ & $1.5 - 4.8$ & $\simeq 9$ & $8$ & $2.2 \pm{0.8}$ & $3.5$ & $^{a}$ \\
        
        IGR J17511-3057	   & $4.1$ & $2009, 2015$ & $0.57 - 4.9$ & $< 32$ & $6.9^{+0.6}_{-2.5}$ & $< 5.8$ & $ < 5 $ & $^{b}$\\

        IGR J17494-3030    & $2.7$ & $2012, 2020$ & $0.009 - 0.77$ & $< 2.3^{*}$ & $\simeq 8$ & $14.5 \pm 5$ & $6.3$     & $^{c}$ \\
                            
        SAX J1808.4-3658   & $2.5$ & $1998, 2000, 2002, 2005$ & $0.041 - 2.7$ & $\simeq 0.5$  & $3.5 \pm 0.1$ & $0.7$ & $1.4$ & $^{d}$\\
                           &       & $2008, 2011, 2015, 2019, 2022$ & & & & & &  \\            
							
       XTE J1751-305      & $2.3$ & $2002, 2005, 2007, 2009$ & $3.4 - 13$   & $\sol 2.3$ & $8.5^{+0.5}_{-1.8}$ & $2.9 \pm{0.6}$ & $2.6$ & $^{e}$\\
        IGR J00291+5934    & $1.7$ & $2004, 2008, 2008, 2015$ & $0.23 - 2$ & $\sim 0.1 - 1$ & $4.2 \pm 0.5$ & $0.9$ & $1.2$ & $^{f}$\\

        \hline
            \end{tabular}
            $(a)$\citealt{Krimm2007DiscoveryCompanion,Markwardt2007SWIFTPulsar,Patruno2010TheJ1756.92508,Sanna2018SWIFTOutburst} ; $(b)$\citealt{Galloway2008BiasesBursts,Markwardt2009RXTEDetected,Altamirano2010TypeJ175113057,Falanga2011SpectralIGRJ175113057,Riggio2011TimingIGRJ17511-3057, Archibald2015, Papitto2016TheXMM-Newton} ; $(c)$ \citealt{Boissay2012, Padilla2013,Chakrabarty2013, Ng2021} ; $(d)$ \citealt{Wijnands1998TheJ1808.43658, Campana2004, Galloway2006, Galloway2008,Hartman2008, Heinke2009,Patruno2017, Bult2020} ; $(e)$ \citealt{Markwardt2002DiscoveryJ1751305,Papitto2008MeasuringJ1751-305,Riggio2011SecularJ1751-305,Archibald2015} ; $(f)$ \citealt{Jonker2005ChandraQuiescence, Markwardt2007OutburstJ1751-305, Papitto2011SpinPulsar, DeFalco2017TheObservations}
	   
\end{table*}

\begin{table*}
	\centering
	\caption{Non-pulsating persistent LMXB sources used in our analysis (see the text for selection criterion). In the Method column, we indicate the methods used for distance estimation (PRE: photospheric radius expansion, GC: Globular Cluster, and PH: Photometry).}

	\label{nonpulsating}
	\begin{tabular}{cccccccc} 
    \hline

 	 Source	    & $P_\mathrm{orb}$ & 	Distance    &$ F_\mathrm{X,avg}$ & $\Mdotin$       & $L_\mathrm{X,avg}$  & Method&   Reference   \\ 
        
	                &	   (h)	       &	   (kpc)	 &	( \ergperscm )    &	( \gpers )       &	( \ergpers )       &      &	              \\
       
        \hline

        4U 1820-30	    & $ 0.19   $ & $ 7.60 \pm 0.40 $ & $ 4.48\x 10^{-8} $ & $  1.67\x 10^{18} $ & $ 3.10\x 10^{38} $ & PRE &$^{\mathrm{(1)}}$ \\
        Ser X-1		    & $ 2.00   $ & $ 6.90 \pm 0.90$ & $ 2.93\x 10^{-8} $ & $  8.98\x 10^{17} $ & $ 1.67\x 10^{38} $ & PRE & $^{\mathrm{(1)}}$ \\
        Cyg X-2		    & $ 236.20 $ & $ 10.25 \pm 1.35$ & $ 1.07\x 10^{-8} $ & $  7.23\x 10^{17} $ & $ 1.34\x 10^{38} $ & PRE & $^{\mathrm{(1)}}$ \\
        GX 17+2		    & $	-	   $ & $ 9.80 \pm 1.30$ & $ 1.03\x 10^{-8} $ & $  6.37\x 10^{17} $ & $ 1.18\x 10^{38} $ & PRE & $^{\mathrm{(1)}}$ \\
        Sco X-1		    & $ 18.90  $ & $ 2.80 \pm 0.30$ & $ 6.75\x 10^{-8} $ & $  3.41\x 10^{17} $ & $ 6.33\x 10^{37} $ & PH & $^{\mathrm{(3)}}$ \\
        SLX 1735-269    & $		   $ & $ 5.15 \pm 0.65$ & $ 1.09\x 10^{-8} $ & $  1.86\x 10^{17} $ & $ 3.45\x 10^{37} $ & PRE & $^{\mathrm{(1)}}$ \\
        4U 1735-44	    & $4.65	   $ & $ 6.16 \pm 1.04$ & $ 6.88\x 10^{-9} $ & $  1.68\x 10^{17} $ & $ 3.12\x 10^{37} $ & PRE & $^{\mathrm{(1)}}$ \\
        1A 1246-588	    & $	-	   $ & $ 3.40 \pm 0.40$ & $ 2.01\x 10^{-8} $ & $  1.49\x 10^{17} $ & $ 2.77\x 10^{37} $ & PRE & $^{\mathrm{(1)}}$ \\
        4U 1812-12	    & $ 1.90   $ & $ 2.65 \pm 0.35$ & $ 2.42\x 10^{-8} $ & $  1.10\x 10^{17} $ & $ 2.04\x 10^{37} $ & PRE & $^{\mathrm{(1)}}$ \\
        GX 3+1		    & $ 18.50  $ & $ 5.10 \pm 0.70$ & $ 5.57\x 10^{-9} $ & $  9.33\x 10^{16} $ & $ 1.73\x 10^{37} $ & PRE & $^{\mathrm{(1)}}$ \\
        4U 1746-37	    & $ 5.16   $ & $ 11.00^{+0.9}_{-0.8}$& $ 8.42\x 10^{-10}$ & $  6.56\x 10^{16} $ & $ 1.22\x 10^{37} $ & GC & $^{\mathrm{(2)}}$ \\
        4U 1728-34	    & $ 1.10   $ & $ 3.85 \pm 0.55$ & $ 5.45\x 10^{-9} $ & $  5.21\x 10^{16} $ & $ 9.67\x 10^{36} $ & PRE & $^{\mathrm{(1)}}$ \\
        4U 1724-307	    & $2,160.00$ & $ 7.40 \pm 0.50$ & $ 1.26\x 10^{-9} $ & $  4.44\x 10^{16} $ & $ 8.26\x 10^{36} $ & PRE & $^{\mathrm{(1)}}$ \\
        XB 1254-690	    & $ 3.93   $ & $ 7.60 \pm 0.80$ & $ 1.16\x 10^{-9} $ & $  4.30\x 10^{16} $ & $ 7.99\x 10^{36} $ & PH & $^{\mathrm{(4)}}$ \\
        4U 0513-40	    & $0.28	   $ & $ 12.10 \pm 0.30$ & $ 1.53\x 10^{-10}$ & $  1.44\x 10^{16} $ & $ 2.68\x 10^{36} $ & GC & $^{\mathrm{(2)}}$ \\
        4U 1705-44	    & $10.00   $ & $ 5.80 \pm 0.80$ & $ 6.11\x 10^{-10}$ & $  1.32\x 10^{16} $ & $ 2.46\x 10^{36} $ & PRE & $^{\mathrm{(1)}}$ \\
        M15 X-2		    & $0.38	   $ & $ 10.30 \pm 0.40$ & $ 1.10\x 10^{-10}$ & $ 7.51\x 10^{15} $ & $ 1.40\x 10^{36} $ & GC & $^{\mathrm{(2)}}$ \\
        4U 1850-087	    & $0.34	   $ & $ 8.20 \pm 0.60 $ & $ 1.54\x 10^{-10}$ & $  6.68\x 10^{15} $ & $ 1.24\x 10^{36} $ & GC & $^{\mathrm{(2)}}$ \\
        GX 349+2	    & $22.50   $ & $ 4.90^{+1.7}_{-2.9}$ & $ 9.53\x 10^{-9} $ & $  6.29\x 10^{15} $ & $ 1.17\x 10^{36} $ & PH & $^{\mathrm{(5)}}$ \\
        XB 1832-330	    & $2.15	   $ & $ 9.60 \pm 0.40 $ & $ 1.03\x 10^{-10}$ & $  6.10\x 10^{15} $ & $ 1.13\x 10^{36} $ & GC & $^{\mathrm{(2)}}$ \\
        AC 211		    & $17.10   $ & $ 10.30 \pm 0.40$ & $ 7.00\x 10^{-11}$ & $  4.78\x 10^{15} $ & $ 8.89\x 10^{35} $ & GC & $^{\mathrm{(2)}}$ \\
        4U 0614+09	    & $0.86	   $ & $ 2.29 \pm 0.30$ & $ 1.10\x 10^{-9} $ & $  3.72\x 10^{15} $ & $ 6.92\x 10^{35} $ & PRE & $^{\mathrm{(1)}}$ \\
        SLX 1737-282    & $11.00   $ & $ 4.50 \pm 0.60$ & $ 1.36\x 10^{-10}$ & $  1.77\x 10^{15} $ & $ 3.28\x 10^{35} $ & PRE & $^{\mathrm{(1)}}$ \\
        2S 0918-549	    & $0.29	   $ & $ 3.45 \pm 0.45$ & $ 1.23\x 10^{-10}$ & $  9.45\x 10^{14} $ & $ 1.76\x 10^{35} $ & PRE & $^{\mathrm{(1)}}$ \\

        \hline  
    \end{tabular}
    
    $(1)$ \citealt{Galloway2020}; $(2)$ \citealt{Kuulkers2003}; $(3)$ \citealt{Bradshaw1999}; $(4)$ \citealt{Gambino2017}; $(5)$  \citealt{Arnason2021} 
     
\end{table*}



\bibliographystyle{mnras}
\bibliography{References.bib} 








\bsp	
\label{lastpage}
\end{document}